# Realism is almost true: A critique of the interface theory of perception

Arman Rezayati Charan[1], Shahriar Gharibzadeh[1,*], S. Mahdi Firouzabadi[1]

[1]Institute for Cognitive and Brain Sciences, Shahid Behehshti University, Tehran, Iran

*Corresponding author, Email: s_gharibzade@sbu.ac.ir

Abstract

Objective reality and its relation to our perception have been an important topic of debate in philosophy and cognitive science. Hoffman's "interface theory of perception," which asserts that our perception has no congruency with reality, is recent and controversial among existing theories.
Hoffman and Prakash formulated and evaluated their theory using evolutionary game theory and genetic algorithms. This paper investigates the "interface theory of perception," introducing an agent-based simulation. Using the principles and hypothesis of Hoffman's model, first, we regenerate and approve his initial claims to consider interface as a winning perception strategy. Then we move forward to assess his model in more realistic conditions and challenge interface perception model. Our findings indicate that in case of drastic environmental changes, interface perception is no longer compatible with reality and pushes the interface species further toward extinction. Our proposed model will pave the road for future studies to investigate the perception strategies in a more realistic condition.

Introduction: The relation between human perception of the world and reality is an old problem in philosophy. Many philosophers have discussed this topic from philosophers of ancient greek like Plato and Aristotle and figures from the age of enlightenment such as John Locke and Immanuel Kant.

We can classify Different theories and opinions on this topic into two general categories: realism and idealism. The realistic view believes an external reality exists. In its extreme version, realism asserts that there is complete congruence between objectivity and subjectivity.  This extreme version is called naive realism  (Fish 2010)   (Nuttall 2013)   (Hoffman and Prakash 2014) .

On the other hand, some philosophers advocate some level of discrepancy between subjectivity and objectivity. Among them is the influential philosopher Immanuel Kant known as an idealist. Kant claimed that we perceive the world in this unique way only because of the specific structure of our minds. According to Kant, there are categories in our minds that classify input signals from the objective stimulus. If we perceive the world in specific ways, it is because of the constraints imposed by those mentioned categories. Based on this theory, there is no reason for the veridicality of our perception (Kant 1998). This debate has continued up to the contemporary age and extended into modern cognitive science.

Although this debate includes perception in its general sense, for practical reasons, most studies, particularly in cognitive science, focus on visual perception. The dominant opinion in classic studies about the vision was that our visual system reflects the world as it is. In recent decades, numerous researches argued that visual perception is practical only when it precisely describes the objective world (Marr 1982) (Palmer 1999). So they conclude that evolution favors the veridical perceptual system above the non-veridical one. For example, an evolutionary biologist says, "Our sense organs have evolved to give us a marvelously detailed and accurate view of the outside world. We see the world in color and 3-D, in motion, texture, non-randomness, embedded patterns, and a great variety of other features, likewise, for hearing and smell. Together the design of our sensory systems blesses us with a detailed and accurate view of reality, exactly as we would expect if the truth about the outside world helps us to navigate it more effectively" (Trivers 2011).

On the other hand, from a very radical point of view, the interface theory of perception claims there is no need for any congruency between the objective reality and our perception. There is no such congruency in current evolved perceptual systems.

D. Hoffman as the author of this theory, argues that evolution and natural selection have shaped our perception of the world in a way to be fit for survival and reproduction. Quite simply, perception is about having kids, not seeking the truth! (Hoffman, Singh et al. 2015).

Based on this theory, our perceptual system does not project the world accurately, and there is no need to do so. To be more precise, there is no kind of homeomorphism between our perception and the real world. The central premise of this theory is that the criterion of evolution for choosing a perceptual system is that system's ability to guarantee the survival of a living species rather than presenting reality as accurately as possible. In addition, Hoffman argues that representing exact reality is a very energy-consuming operation. The more accurate the representation, the more energy consumed.

They propose the "interface theory of perception" as a theoretical framework for non-veridical perception, offering less energy consumption and a better chance of survival.

The Windows desktop metaphor is a convenient way to describe the central concept of the Interface Theory. When someone selects an icon on a Windows desktop and removes it by dragging it to a recycle bin, in reality, there is neither an icon nor a recycle bin on the computer. In computers, there are circuits, diodes, and many other electronic devices. The desktop interface hides the complexity of computer hardware in the same way that the perceptual interface would hide the complexity of reality and instead provides a specific guide to action. Based on this metaphor, the perceptual system that behaves like an interface and guarantees survival less expensively would be more optimized. (Mark, Marion et al. 2010; Hoffman and Prakash 2014; Hoffman, Singh et al. 2015, Hoffman 2016).

In this paper, by inspiring the key assumptions introduced by D. D. Hoffman (Mark, Marion et al. 2010), an agent-based model has been created to investigate the "Interface theory of perception." They introduced three species representing three different perceptual strategies based on naive realism, critical realism, and interface theory. Then designing an evolutionary game between these three strategies, they show that interface is the stable evolutionary strategy. We employ similar configurations of the evolutionary game between these three species and set up an agent-based simulation to investigate which one outcompetes the others. Similar to Hoffman's claim, our results approve that interface strategies prevail when the agents do not face drastic environmental changes. It means that in a specific environment in which almost all essential features are constant, interface agents may do better than any other species. However, when dramatic changes happen in the environment, the current perceptual interface serves this species's detriment and will no longer remain compatible. So due to this incompatibility, interface agents are more likely to extinct. (Mayr 2001)

The current study will show that our perception might not be the same as external reality. However, contrary to Hoffman's theory, it is not in conflict with reality but rather an approximation of external reality. Some experimental studies also confirm this view. One experiment that fits well with this view examines the ability of people to estimate the age of people of the opposite sex based on their voices. In these experiments, for example, women will listen to the audio files of men of different ages and then be asked to identify the age of men. The results indicated that the age detected by the test subjects has a high level of correlation with the actual age of speakers.

Meanwhile, the sexual utility function is bell-shaped in terms of reproductive characteristics, with young adults ,based on their reproductive capability, on the pick and very young and older people having equally low reproductive utility at the tail ends of the curve. Thus, according to Hoffman's theory, people should not distinguish the voice of children and the elderly of their opposite sex. Nevertheless, both scientific experiments and our everyday experience prove this to be wrong. For example, in a study reported in (Zäske and Schweinberger 2011) subjects from both male and female gender were asked to listen to recorded voices of people of the opposite sex and use them to estimate the speaker's age. The results showed that the estimated age had a very high positive correlation with the actual age of the speakers. Another similar study found that although children and the elderly have a more accurate estimate of the age of speakers of the opposite sex, everyone can distinguish between different age ranges with acceptable accuracy, which means that they all have a good estimate of reality.(Hughes and Rhodes 2010)
.

# Investigating Interface Theory Using Agent Based Models

## Agent-based models

Agent-based models (ABM) are made up of autonomous decision-making entities called agents. The interaction of these artificial agents with the environment and other agents creates a dynamic system (Bonabeau 2002, Epstein 2006) . Agent-based modeling is a paradigm to model and simulate social interactions with a bottom-up approach. Every agent-based model has some essential components like agents, environment, the topology of interaction, the technology of decision making, and learning. Simulating the behaviors and interactions of fundamental elements, ABM brings an opportunity to capture the emergent phenomena and nonlinear behaviors in the system.

The environment of an ABM is the set of background conditions in which agents locally or globally are engaged with them. We can define any topology for the interactions of agents. For instance, in the current model, the environment is a 2D surface composed of patches where every agent can occupy a patch and move freely from one patch to another. Also, we can use learning facilities for agents. For example, agents could learn from their experiences based on different machine learning or statistical methods. The other important component of an agent-based model is the rules of decision-making. Different methods are applied to determine how agents make decisions. For example, agents can use the rules of game theory, follow some conditional rules in the form of "if-then," or make decisions according to a specific mathematical equation.

## Simulation setup

Following (Mark, Marion, et al. 2010) definition of agents, we introduce three species of agents: truth-agents, simple-agents, and interface-agents, where these three species respectively represent naïve realist, critical realist, and interface perception systems. The initial energy level of agents and the resource value of patches are assigned randomly from a normal distribution. Each agent possesses a level of energy that they need to preserve above a certain level to survive. Time steps in agent-based modeling are discrete. At each step, we update the status of the agents and environment by applying the rules that dictate the behavior of agents, patches, and their interactions.

In the current model, at each time step, agents evaluate the resources of the neighboring patches and moves toward the patch perceived with the highest utility. It is allowed for the agents to step on the same patch with other agents. For this research ABM simulation, we have used NetLogo (Wilensky 1999). Netlogo environment topologically is a torus, so agents never leave the environment until they die.

## Agents setup

There are three species or breeds of agents with different perception strategies: simple, truth, and interface agents. The initial energy of those agents guarantees an agent's life until reaching the death threshold. After the initiation, agents begin to move around the environment, searching for food resources to earn the energy required for their survival. At the same time, agents' internal perceptual systems spend energy processing information and perceiving the outside world. If agents come below a specified threshold of energy while perceiving the outside world searching for food, they will die.

**Truth Agent** perceives the world as it is. As explained earlier, we assign the food value to each patch from a specific normal distribution. In theory, each patch can adapt value from an infinite set of values, but we divided the food range into 100 categories for convenience and simplicity. So truth agents have 100 perceptual categories that indicate discrete values for the amount of food (resource) available in each patch. Truth agents' perception coincides with the objective reality; thus, they percept food resources by their actual quantities. The perception function of the truth agents, which we will show as $p_T(.)$ is the identity function.

**Simple agents** are not as exact as the truth agents in their perception, but their perception has some congruency with reality. Their characteristic feature is that their perceptual system makes an approximation of reality. In our model, for any simple agent, there are two thresholds, one upper and one lower, which divide the values of food resources into three categories. In other words, the simple agent cannot specify the exact quantity of food, but instead, they can distinguish between low, medium, and high quantities and quantities of food resources. There is a homomorphism between their perceptual categories and the reality, i.e., if a and b are two food resources with values $v_a$ and $v_b$ the simple agents perceive them as $p_s(v_a)$, and $p_s(v_b)$ respectively, then we have:

$$if\ v_a > v_b\ then\ p(v_a) \geq p(v_b) \quad (1)$$

In other words perception function of simple agents ($p_s(.)$) preserves the standard order of real lines.

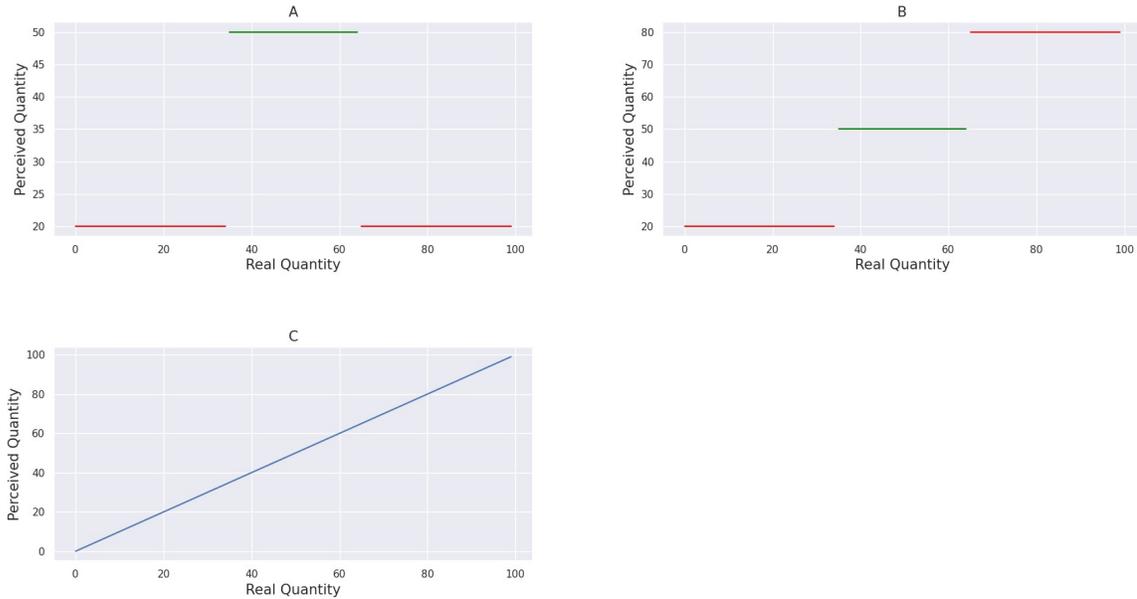

*Figure 1:A. Perception function of an interface agent when its interface is compatible with the environment.*
*B. Perception function of a simple agent.*
*C. Perception function of a truth agent.*
*Green lines show the areas with highest utility and red ones show areas with lowest utility.*

### Interface Agents:
The interface agent has developed a perception system in which perceptual categories align with the utility function and not the reality.
For instance, when interface agents find two resources with different amounts of food, the resource that provides more utility is perceived with greater food quantity. In other words, the interface perceptual system perceives the world in a way to guarantee to get higher utility.
For simplicity and ease of comparison, similar to (Mark, Marion, et al. 2010), we suppose that the utility function by food resource is Gaussian for all agents:

$$U(x) = a e^{\frac{-(x-b)^2}{2c^2}} \quad (2)$$

Eq. (2) is the general form of a Gaussian function in which parameters a and b are arbitrary constants. This function form is adapted from (Mark et al., 2010) as we intend to use the same setup to investigate their key findings.
For interface agents, a threshold in utility separates all food resources into two perceptual categories of high and low-quality food resources. Fig. (1) indicates that the interface agents find the high utility in green color and those with low utility red. The range of food resource value is divided into three, similar to simple agents. Fig. (1) indicates the alignment between utility and perception cause that interface agents perceive resources with different real food values in the same perceptual category.

### Agent Based Model Parameters

The simulation begins with creating agents and the environment, as demonstrated in Fig.(2). We assign agents with the random initial energy level and patches with random resource value. The overall distribution of resources and energy follows a normal distribution with specific parameters. Also, there are other parameters like "food growth rate," "cost per bit of information," "boundary on food," "environment change probability". Changing the game's parameters could affect the outcome in a manner that changes the fate of each species. Below we discuss the parameters in detail:

**The food growth rate** is the rate of resource renewal in each step of the simulation. The higher food growth rate would lead to a more resourceful environment, thus the higher chance for the species to retrieve their energy level and survive.

**Cost per bit of information**: This is a crucial parameter in interface theory. It will define how much energy is consumed by each species to assess and choose from available resources. Also, it will determine the cost of possessing knowledge about the utility of food amounts for each agent. In our model, similar to (Mark, Marion, et al. 2010), we assume all agents can store the same amount of energy. Clearly, truth agents consume more energy than other agents because they have much more perceptual categories. The other worth mentioning point here is that the energy cost of acquiring each unit of knowledge is identical for all kinds of agents because they have the same utility function.

**The boundary on food**: This parameter is specific to simple and interface agents and determines the width of the food value area that the interface perceives as green (high-quality food) also it determines the different quantity categories for simple agents.

At each step of the simulation, each agent first evaluates the patches in its neighborhood according to its perceptual system and then selects the food resource with the highest utility among them. Since the truth agent knows the exact amount of each resource and its corresponding utility, it selects precisely the best of the available resources. The simple agent recognizes only three food categories and can distinguish between low, medium, and high. The simple agent knows that low and high quantity resources are equally desirable, and resources with medium quantity are more desirable than the other two categories. The interface agent only knows that there are desirable and undesirable resources and perceives all desirable resources with the same quantity, and more importantly, he sees the quantity of all undesirable resources as the same. The "boundary on food" variable defines the boundaries between the desired region and the undesirable regions. The difference between simple agents and interface agents is in the number of quantity categories. The interface agents perceive the first and last regions in fig.(1A) with similar quantity because they have the same utility and as we mentioned above, utility is equivalent to quantity for the interface agents.

The energy consumed in each step of simulation for each species is (Mark et al., 2010):

$$C = c_e q + c_q q n_b \quad (3)$$

We set $q=100$ for the truth agents, $q=3$ for the simple agents, and $q=2$ for the interface agents in which "$q$" is the number of perceptual categories for each agent. $c_e$ denotes the cost per bit of information and $c_k$ denotes cost per bit of knowledge about utilities. Following the assumptions by Hoffman, we set $c_k = \frac{c_e}{10}$. $n_b$ is the number of bits that are used to represent the utility of a food resource which similar to (Mark, Marion et al. 2010)In this simulation, to avoid unnecessary complications, we have considered only one type of food resource with normal distribution over the environment. Finally, at each step of the simulation, if agents choose any food resource with quantity "x," then the gain would be as the following:

$$Gain = U(x) - C \quad (4)$$

Which U is the utility of resource x, and "C" is the cost of perception that will be determined based on the agent's type.

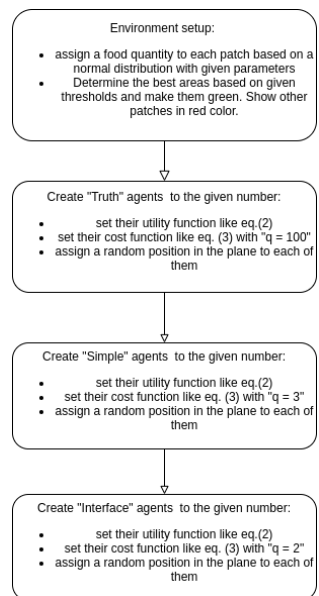

*Figure 2: Procedure of agents and environment setup*

## Models with environmental change

The critical assumption of interface theory is that evolution drives all living systems to align their perception system to the world so that less energy consumption can guarantee their survival. So we can go further and ask, what if the environment changes dramatically? What would happen for species with an interface perception system?

**Environment change probability** is a parameter that sets the possibility of environmental changes in each step. This assumption would get us closer to a more realistic model. In other words, for species with a veridical perception system, the perception is exact and independent of the environment, so such living systems are more likely to survive facing drastic environmental changes. However, interface perception is compatible with the current environment, which will be useless as soon as the environment changes. Hoffman has mentioned (Hoffman, Singh et al. 2015) many examples in which environmental change has led to false perceptions, thus endangering the lives of many species. For instance, some beetles mistake shiny trash bottles left in nature as an eligible candidate for mating, which has disturbed their reproduction process. Hoffman mentioned these examples as evidence to confirm the interface theory of perception, but in contrast, however, these phenomena are well explicable within the framework of interface perception. They also highlight the weakness of the interface perceptual system. They indicate that species with the interface perceptual systems are prone to extinction in case of incompatibility between perceptual interface and the new environmental reality. We argue that the interface perception system is not the best strategy from the evolutionary point of view and investigate this claim using an agent-based modeling approach.

After any drastic change in the environment, the species with an interface built upon the previous state of the world will not be able to perceive the world effectively. On the other hand, the course of natural evolution that occurs slowly over a long period can not help the interface species adapt to the change before extinction.

Therefore, we can consider the examples provided by Hoffman as local optimums. Interface species will not survive unless they can build a mechanism, yet unknown, to adjust perceptual interface according to the changes of the environment.

Thus, based on this consideration, facing environmental changes, the utility function of "Truth" and "Simple" remains unchanged while the interfaces' perceived utility will change as described in the fig. (3).

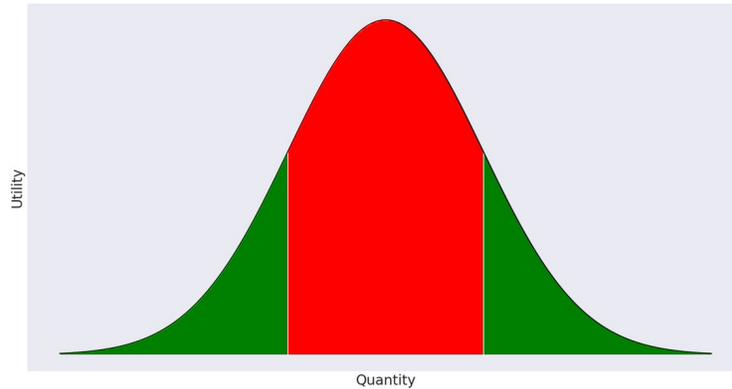

Figure 3: Perception function of interface agent after drastic change in the environment.

Green areas are where the interface agent (mistakenly) perceive them as high-utility resources, and red areas where perceive them as low-utility. Based on this consideration, the interface's energy at step n of the simulation will be as follows:

$$E^n_{interface} = E^{(n-1)}_{interface} + (1-p)^{(n-1)}(pU^{\square}) + (1-p)^n U - nC \quad (4)$$

with the following parameters:

$E^n_{interface}$: Energy of each interface agent in step n
$p$: probability of dramatic change in the environment
$U$: utility gained by choosing the food based on compatible interface perception function

$U^\square$: utility gained by choosing the food based on reverse interface perception function
$C$: Cost, according to Eq. 3

In the models with environmental change, the environment and agent's first setup is like Fig.(2). After initial setup, the simulation begins, and in each step, the state of agents and environment will change. The following diagram will explain the sequence of events in the simulation.

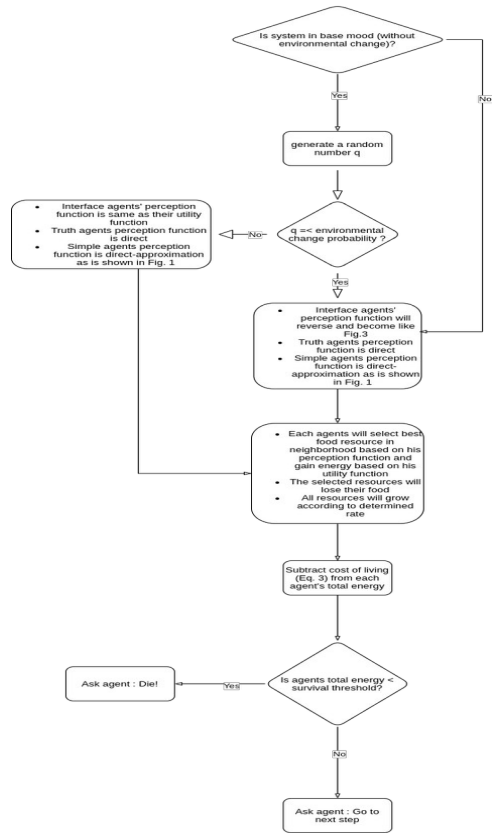

*Figure 4: Simulation procedure at each step.*

# Results

The results will show how the population of each species would change during the simulation by changing values of different parameters, including "cost per bit of information," "food growth rate," "environment change probability," and "boundary on food."

In order to create a fair competition to assess species survival capability, we initiated all simulations with 2000 agents from each species in base mode and 1000 agents in the environment changing mode. 'food growth rates as an indicator for environment resource renewal rate in each step has changed from 10 to 60 and 70 percent respectively for environment changing and base modes. As the key influential parameter of the model, cost per bit of information ranged between 0.01 to 1 and 1.5 percent of energy capacity respectively for base and environment changing modes. Also, to check dependency for different values of 'boundary on food,' we have investigated the results setting this parameter in a range of 10 to near 50. First, we review results without considering environmental changes, which corroborates the earlier claims of interface theory by D. Hoffman. Afterward, we further investigate the model, considering the environment change as a new parameter in simulations to challenge claims by Hoffman.

## Results without environment change

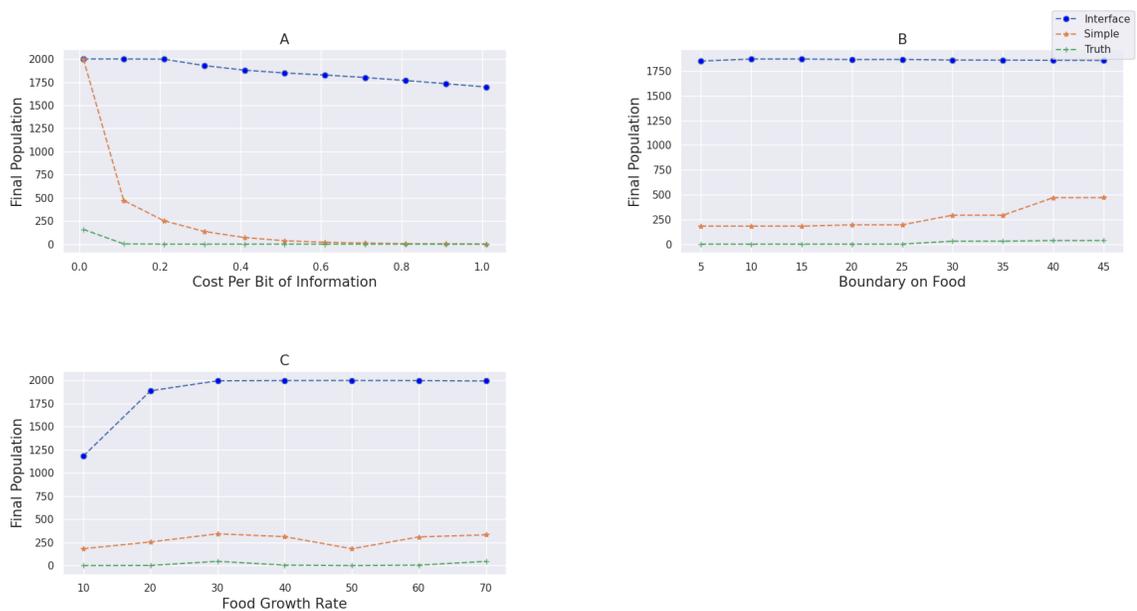

*Figure 5: Average final population of 3 different species respect to: A. Cost per bit of information, B. Boundary on food, and C. Cost per bit of information*

To compare the survival chance of species for every set of initial parameters, we run the simulation 30 times. The average number of survivors of each species at the end of the simulations would indicate the chance of survival compared to other species. Figure (5A), shows the final average number of each species according to "cost per bit of information."
For the lower values of cost per bit of information, all species continue to survive; however, the interface species are more abundant. As the energy cost of information increases, the final number of the simple strategy quickly converges to zero, like the truth's population. However, the population of interfaces experiences a lower decrease. In other words, by increasing the cost per bit of information, interfaces dominate and drive other species to extinction.
In figure (5B), we see the average of survivors of 3 different species respect to the "boundary on food". Like the previous case, the interfaces are the dominant species, and truth agents go to distinction. The same occurs in the case of food growth rate which is shown in figure (5C).

## Interface agents facing environment change

In this case, we investigate the effect of random environmental changes on different species' chances of survival. If the probability of change is not zero, we will apply a probabilistic radical change within the simulation environment. This change leads to an incompatibility between the interface agent's perception system and the environment. In other words, the interface agent's perception of the world is no longer aligned with its utility function. Therefore the previously adapted "interface" does not benefit interface agents but will push the interface further toward extinction. This case is well understood within the Desktop metaphor when malware or virus infects the computer system; it changes the functionality of predefined interfaces.

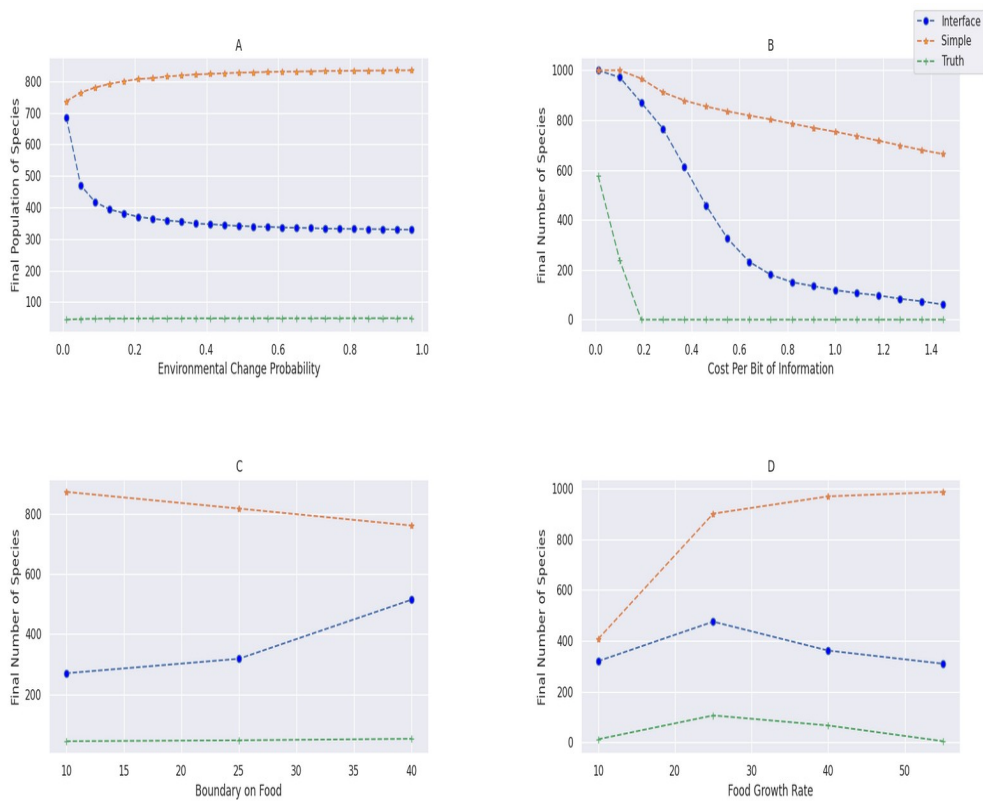

*Figure 6: Average final population of different strategies in case of environmental change respect to: A. Environment's change probability, B. Cost per bit of information, C. Boundary on food, D. Food growth rate.*

Figure 6 shows the simulation results in the environment which can change drastically. As can be seen, for all parameters, the simple strategy performs significantly better than the interface strategy. In this case, too, the truth strategy has no chance of survival. As shown in Figure (6A), even at small probabilities of environmental change, the performance of the simple strategy is much better than the performance of the interface strategy. However, the interface strategy is not entirely extinct due to its low energy consumption and the minimal number of perceptual categories. Figure (6B) shows the population changes of different strategies regarding changes in the "cost per bit of information" value. Here, as the cost increases, the population of all strategies decreases, but still, the simple strategy performs much better than the interface, and even at high-cost levels, the majority of its population will survive while the interface population goes extinct. Figure 6c shows the population changes of different species compared to the "boundary on food" changes. Also, here, the simple agent outperforms the interface kind, but the interesting point is that, unlike in the case where no change in the environment occurs, the interface performance improves by increasing the boundary. Here, the interface perception function works misleadingly and sees the points with high utility as low utility points, and naturally, the narrower the best area range, the interface suffers less from a flawed perception system.

In Figure (6D), increasing the "food growth rate" above a specific point (around 25) would decrease the interface's chance for survival. The reason is that the higher value of food growth rate guarantees that all the species would constantly access resources with high food quantity. However, remember that high quantity food is indeed less valuable to the interface because of the interface utility function. The interface truly considers a giant whale a less valuable food resource than a tuna fish.

Due to the change in the environment, the interface perception misleads the species into finding less desirable high quantity food resources as desirable resources, and therefore by eating from them, the interface gains less energy and will have worse performance.

Below, we can see the heat map of results, which shows the final population of different kinds of agents regarding "environment change probability" and "cost per bit of information, " which are the most critical variables among all. As can be seen, the simple species can survive in almost all of the parametric space, while the interface species can only survive if both the "cost per bit of information" and the probability of changing the environment are low at the same time. Even in cases where the "cost per bit of information" is low but the probability of environmental change is high, the interface population is close to zero in many runs of the simulation.

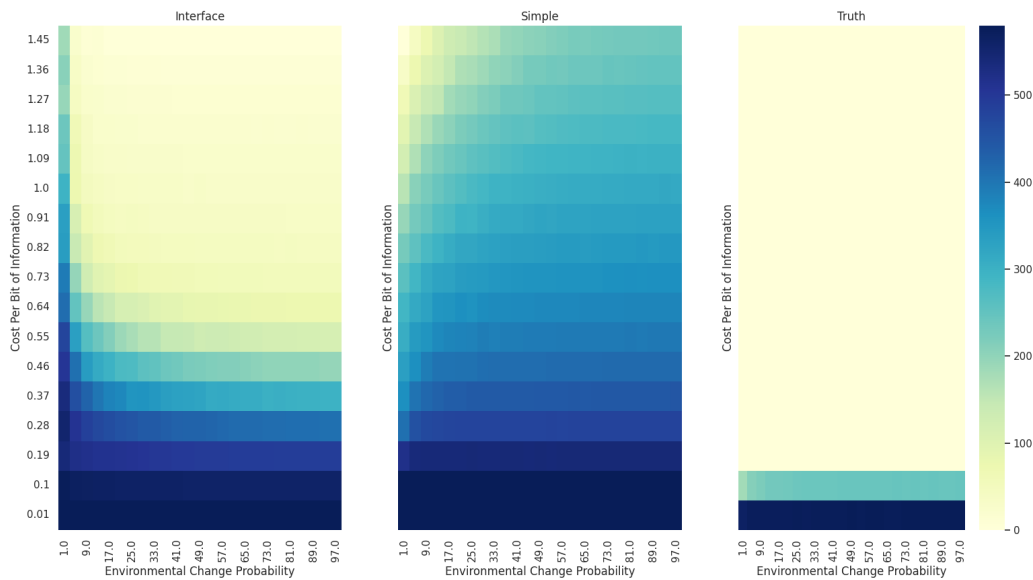

Figure 7: Final population of different strategies due to different values of "cost per bit of information" and "Environmental Change Probability".

# Discussion

In this paper, first, using a similar setup by Mark, we build an agent-based model to investigate the interface agents and reproduced the same results. (Mark, Marion, et al. 2010). We evaluate the models with three parameters: "boundary on food", "cost per bit of information" and "food growth rate".

The first parameter determines the perceptual categories of the interface and simple agents. It will define the ranges of food value that interface agents perceive as high-quality green or low-quality red food. In our model, increasing the value of "boundary on food" for interface agents makes green resources rare but with higher utility. The second parameter determines the cost of perception precision or, in other words, the cost of energy for agents' save knowledge about their utility function and acquire information about the utility of the environment. Obviously, this cost suffers the truth agents much more than any other agents due to the high number of perception categories. As (Mark, Marion, et al. 2010) results have shown, interface agents drive truth agents to extinction in a large area of both parameters. So they concluded that non-veridical perception compared to veridical perception would count as an evolutionary advantage.

Here, to create a more realistic model, we considered two additional parameters in our models: "environment change probability" and "resource substitution or renewal rate." Each agent probes the environment and finds energy resources according to its decision-making process and perceptual strategy. The first part of our simulation results confirms the interface theory's predictions similarly to Mark (Mark, Marion, et al. 2010). In the whole parameter space, interface agents significantly outperform the other ones(Fig.5). The truth agents in almost all situations vanish, and the simple ones in some areas of the parameter space survive and coexist with interface species. So we can conclude that if there were no additional constraints like environmental change, our models and simulation confirms Hoffman and colleagues' results.

In the second part of our results, in contrast to the claims of interface theory of perception, the interface agents are not preferred by evolution, and the simple agents have the evolutionary advantage. As we can see in figs 6 and 7, the simple agents had a better chance of survival and outperformed interface agents for almost all parameter space. As we defined before, simple agents are somewhere in between interface and truth strategy. Simple agents are not idealists,100% loyal to reality like truth agents. Nor are they like interface agents, complete pragmatists, caring for utility and with total disregard to reality. The simple agents' homomorphism between their perception and reality makes them better equipped to adopt environmental changes and survive longer than other species. It is important to note that similar to the results by Mark (Mark, Marion, et al. 2010) , in both our cases, the truth agents have no chance of survival which confirms that veridicality is no evolutionary advantage at all.

As can be seen from Figures 6 and 7, the simple species performs much better than the interface one, and this superior performance is visible for different values of "probability of environmental change," "cost per bit of information," and "food growth rate." The difference in performance between the two strategies is in such a way that in all cases, the majority of the simple species population can survive, while the interface species population decreases sharply, and in some cases such as "food growth rate" and "cost per bit of information" this species is becoming extinct. Therefore, our modeling and simulation results show that the simple perceptual strategy has an undisputed advantage over other perceptual strategies if the models take environmental changes into account.

Thus we can claim that most evolved beings have a perceptual system that provides them with an approximation of objective reality. In other words, we as evolved beings can be almost certain that what we perceive is, to some extent, approximately what exists in the outside world.

To further investigations of the above-mentioned perceptual strategies, we can study the effects of new parameters to create a more realistic model and focus on energy consumption and finding food resources. For instance, other than the perception of the environment to find resources, perception of danger and possible threats is highly related to perception models and have great influence on the chance of survival. The learning capacity of species is another significant factor that previous researches have failed to investigate. As observed in nature, the learning capacity creates an effective tool for different species to better adapt to the changes in the environment. We will hope to improve our model by considering learning strategies for species. Several experimental visual and auditory perception studies suggest possible explanations based on different perceptual strategies discussed in this paper. The authors will hope to proceed with current research in the future with the following mentioned ideas.